# UNFAIR LEARNING: GENAI EXCEPTIONALISM AND COPYRIGHT LAW

David Atkinson[1,2]

**_Abstract_**

This paper examines the claim that generative artificial intelligence (GenAI) should be exempt from copyright law when reproducing copyrighted works without authorization, based on a fair use defense. It examines fair use legal arguments and eight distinct substantive arguments, contending that every legal and substantive argument favoring fair use for GenAI applies equally, if not more so, to humans. Therefore, granting GenAI exceptional privileges in this domain is legally and logically inconsistent with withholding broad fair use exemptions from individual humans. It would mean no human would need to pay for virtually any copyrighted work again. The solution is to take a circumspect view of any fair use claim for mass copyright reproduction by any entity and focus on the first principles of whether permitting such exceptionalism for GenAI promotes science and the arts.

---

[1] Assistant Professor of Instruction, Business, Government and Society Department, McCombs School of Business, University of Texas, Austin

[2] A special thanks to Chandler Lawn, UT-Austin law student, for her valuable contributions

# TABLE OF CONTENTS

## I. Introduction

It is no secret that generative AI (GenAI)[3] requires trillions of words and sub-word components (called tokens) to train the most capable models. These trillions of tokens overwhelmingly originate from billions of publicly accessible web pages, the vast majority of which contain content protected by copyright law, including books, songs, videos, and other media. GenAI companies routinely copy this content without authorization, directly forming it into training datasets used to train GenAI models, or they acquire such datasets from third parties, such as Common Crawl.

The legal landscape surrounding this practice hinges on a single, high-stakes question: whether fair use can justify the unauthorized reproduction of copyrighted works on this unprecedented scale. GenAI companies almost certainly committed prima facie copyright infringement by copying billions of web pages, but they argue that their subsequent transformation of these works—and their lack of direct market competition with the source material—renders their conduct lawful under the fair use doctrine. Whether the reproductions, distributions, and derivatives at issue qualify as fair use is literally a billions-of-dollars question.[4] Losing a fair use argument could be an existential threat to most GenAI companies.

---

[3] GenAI refers to AI that uses training data to create new content, such as text, images, videos, and audio. For the purposes of this paper, the terms GenAI, large language models (LLMs), and models are used interchangeably.

[4] *E.g.*, The *New York Times* alleges that OpenAI infringed on 10 million works. The New York Times Company v. Microsoft Corporation, 1:23-cv-11195, (S.D.N.Y.) Under current statutory damages provisions, the penalty for willful infringement is up to $150,000 per work. If the *Times* were to prevail, the total penalty could theoretically reach 10,000,000 x $150,000 = $1,500,000,000,000 ($1.5 trillion). By comparison, OpenAI is currently valued at $300 billion. OpenAI, *New funding to build towards AGI*, OPENAI (Mar. 31, 2025) https://openai.com/index/march-funding-updates/

Scholars have also argued that, as a matter of policy, because the volume of data required to train GenAI models is so large that it can only be accumulated by mass scraping and properly licensing all of the content is allegedly impractical, society should essentially exempt GenAI from copyright law.[5] This, I posit, is a prime example of GenAI Exceptionalism, which refers to the situation where GenAI companies seek rights that either (a) society does not allow for humans or (b) society grants only to humans.[6]

This paper is skeptical of such arguments but ultimately does not take a position on whether such a defense is likely to succeed in litigation. Instead, it asks a more fundamental question: why should the same fair use argument that applies to GenAI not apply at least as well to individual humans using the same types but smaller volumes of copyrighted work? For example, suppose courts permit GenAI companies to download and use virtually any copyrighted works they please without authorization, including novels, movies, and songs. Why not allow humans to do the same? If the logic holds for machines, it should hold even more strongly for humans. In essence, accepting the fair use argument for GenAI would mean that humans should rarely have to pay to acquire any copyrighted works.

This reductio ad absurdum reveals the flaw in broad GenAI exceptionalism. Because the legal arguments under the four-factor fair use defense for GenAI are not materially distinguishable from those any individual human could make, the question must turn to substantive arguments about whether courts should treat GenAI and humans differently. This paper argues that courts should. And, contrary to some legal scholarship, the balance should tip in favor of humans, not machines. Scholars and practitioners too often contort themselves into semantic- and fiction-based pretzels to provide special accommodations for GenAI.[7] Rather than recognizing GenAI as a mere tool, they try to expand and stretch copyright law to grant GenAI exceptional privileges. In contrast, this paper argues that presumptive AI exceptionalism is both bad policy and contrary to the constitutional foundations of copyright law.

---

[5] *See, e.g.,* Mark A. Lemley and Bryan Casey, *Fair Learning*, 99 TEX. L. REV. (2020)

[6] One example would be when GenAI companies claim GenAI outputs are speech protected by the First Amendment even though such speech rights have only ever applied to humans, with even corporate speech cases being based on the speech of the humans that make up the corporations. Another example would be when GenAI companies argue that their scraping does not violate terms of service when their bots visit and make a copy of the terms even though terms are generally enforceable against humans who visit the terms page.

[7] *See* David Atkinson, Jena Hwang, and Jacob Morrison, *Intentionally Unintentional: GenAI Exceptionalism and the First Amendment*, 23 First Amend. L. Rev. 173 (2025).

### A. Scope

This analysis focuses on the for-profit use of copyrighted works in creating general-purpose GenAI, such as ChatGPT, Gemini, Claude, and Grok.[8] To be clear, this paper does *not* argue that *no* use of copyrighted works for GenAI may seek refuge behind fair use, but it does take a firm stand that fair use should not be used as a broad and blunt instrument to excuse virtually *all* use of copyrighted works. This paper should also not be misconstrued as saying that no unauthorized use of large databases mainly consisting of copyrighted works can qualify as fair use.[9] Rather, it should be read as arguing that large for-profit companies have no greater claim to fair use arguments than any individual seeking to use the copyrighted work primarily for their own benefit.

Furthermore, when this paper refers to "unfair" learning, it does not mean the fair use doctrine itself is unfair. It means that the application of the fair use doctrine in a way that gives for-profit, oligopoly-leaning tech companies outsized advantages compared to humans is unfair.[10] Such an application is also shortsighted and antithetical to the U.S. Constitution. The Constitution empowers Congress "[t]o promote the Progress of Science and useful Arts."[11] We must presume that Congress passed the Copyright Act pursuant to this enumerated power. Therefore, we should avoid interpreting the Copyright Act in ways that would support outcomes that would undermine the promotion of science and the useful arts. Yet that is arguably precisely what GenAI entities and some eminent legal scholars propose.

---

[8] This paper does not focus on specialized models like those that predict proteins. Additionally, the argument around fair use for GenAI research requires greater nuance, so I set it aside for now. At a high level, some great reasons for leniency for nonprofit researchers include that nonprofits, universities, and research institutions often serve essential public functions, such as providing education, conducting scientific research, and promoting social welfare as their primary roles. Additionally, these organizations often operate on tight budgets and may struggle to afford licensing fees for copyrighted material. Still, for fair use to apply to nonprofit research, it should probably meet at least three criteria: (i) must be released under a non-commercial license, (ii) must be released under a license that allows for scientific use and research only, and (iii) must gate access to the data and models (verify identities of people who want to download the models/datasets), such as with ROOTS and BigCode PII training dataset. Otherwise, it will almost certainly enable and implicitly encourage data laundering.

[9] In addition, this paper also does not apply to the use of public domain or properly licensed works. It *does* apply to "publicly accessible," "publicly available," and "permissively licensed" works, though, as discussed in Section VII(A)(2) *infra*.

[10] It is beyond the scope of this paper to provide a deep dive into how and why GenAI companies have inherent oligopolistic tendencies, but network effects and economies of scale play key roles. For a thorough report on the issue, *see* Jai Vipra and Anton Korinek, Market Concentration Implications of Foundation Models: The Invisible Hand of ChatGPT, THE BROOKINGS INSTITUTION (September 2023) https://www.brookings.edu/wp-content/uploads/2023/09/Market-concentration-implications-of-foundation-models-FINAL-1.pdf

[11] U.S. Const. art. 1, § 8, cl. 8.

Finally, this paper will focus primarily on arguments surrounding data copying (the input side of the equation), not on the models' outputs. However, it is worth noting that even an output that looks unlike any particular input does not automatically make the use of the copyrighted work fair use.

By highlighting the comparison between GenAI and humans, this paper brings fair use arguments into sharp focus, allowing us to consider the enormous implications of notable legal arguments currently unfolding across dozens of high-profile and potentially financially catastrophic lawsuits against GenAI companies and to explore what these prominent legal and policy analyses imply about who in society deserves legal leniency and privileges.[12]

## II. A Brief Overview of How GenAI is Developed

Understanding how GenAI models are created is essential to evaluating fair use claims. For simplicity, this analysis will focus on text processing.

First, GenAI entities collect data in formats such as HTML, PDFs, or images of web pages by scraping the web. This means they deploy "bots" (specialized code) that download some or all content on web pages. Next, the plain text (the words) is extracted from those files. The words are then fed to a tokenizer model, which breaks them down into tokens, consisting of words and subwords (e.g., the word "illegal" might be broken down into "il" and "legal"). These tokens are assigned a token ID (a numerical representation of the token). The token IDs are then algorithmically fed into the model during "pre-training," which adjusts the model weights and creates vectors (long, inscrutable lists of numbers representing the associations between tokens).

This means that every single word from copyrighted sources is fed into the models. Many entire sentences and sequential paragraphs are "memorized" (i.e., copied) within the model verbatim, though it is still unclear exactly how much of a given corpus or document used as training material is ultimately memorized. Importantly, some memorization is a feature, not a bug. A model that memorizes nothing, including no quotes or facts, would not be as useful. However, models cannot selectively memorize *only* facts because they have no inherent awareness of what is fact and what isn't, or what is protected by copyright law and what is in the public domain.

[12] Copyright (direct, vicarious, and contributory) and Digital Millennium Copyright Act claims are by far the four most common claims of the dozens that have been filed. *See* David Atkinson, *A Legal Risk Taxonomy for Generative Artificial Intelligence* (unpublished manuscript) https://arxiv.org/pdf/2404.09479

Regardless of the technicalities, common sense dictates that when a model can "regurgitate" text verbatim, it most likely contains a copy of the training material. GenAI entities try to prevent such regurgitations through several mechanisms, including fine-tuning, in-context learning, reinforcement learning from human feedback, and applying various filters. The very existence of these preventive measures is telling. It strongly suggests that (*i*) these entities almost certainly know they trained the models with copyrighted works, and (*ii*) copies of at least some of those copyrighted works remain embedded within the models.[13]

Several important implications flow from this technical reality. First, these mitigation efforts mean that a "base" model (one without the post-training mechanisms) is more likely to regurgitate content than an "instructed" model (one designed to "chat"). Second, the absence of visible regurgitation does not prove the absence of copying. Third, as Jacqueline Charlesworth, the former general counsel of the U.S. Copyright Office, astutely notes regarding models being prompted to generate an output in the style of a particular artist, "even if the output itself does not rise to the level of infringement, in order to generate recognizable riffs on the artist's works, the model presumably trained on and contains encoded representations of those works. The generation of content in a recognizable style thus points to earlier acts of infringement."[14]

## III. A Brief Overview of Copyright Law

To evaluate fair use claims in the GenAI context, we must first understand the fundamental structure of copyright law. Copyright is a form of intellectual property (others include patents, trademarks, and trade secrets) that grants authors of original works certain exclusive rights immediately upon fixing the work in a tangible medium of expression (*e.g.*, in writing or on a recording).

The exclusive rights include:

- to reproduce the copyrighted work in copies or phonorecords;
- to prepare derivative works based upon the copyrighted work;

---

[13] *See* Shayne Longpre et al., *The Responsible Foundation Model Development Cheatsheet: A Review of Tools & Resources* (2024) at 19 ("For Example, OpenAI's DALL-E 3 model included instructions to avoid outputting copyrighted characters."), https://arxiv.org/abs/2406.16746.

[14] Charlesworth, Jacqueline, *Generative AI's Illusory Case for Fair Use* (August 13, 2024). 27 VANDERBILT JOURNAL OF ENTERTAINMENT AND TECHNOLOGY LAW (forthcoming 2025), at 18, Available at SSRN: https://ssrn.com/abstract=4924997 or http://dx.doi.org/10.2139/ssrn.4924997

- to distribute copies or phonorecords of the copyrighted work to the public by sale or other transfer of ownership, or by rental, lease, or lending;
- in the case of literary, musical, dramatic, and choreographic works, pantomimes, and motion pictures and other audiovisual works, to perform the copyrighted work publicly;
- in the case of literary, musical, dramatic, and choreographic works, pantomimes, and pictorial, graphic, or sculptural works, including the individual images of a motion picture or other audiovisual work, to display the copyrighted work publicly; and
- in the case of sound recordings, to perform the copyrighted work publicly by means of a digital audio transmission.[15]

These provisions make it clear that almost all internet content is automatically copyrighted upon posting to a webpage, which means the exclusive rights protect the content. When a GenAI company scrapes a web page without authorization, it directly reproduces the work (by making a copy). It may also create a derivative work by training models on the reproduced work (if models are essentially a compression of the training data). There may also be a distribution if the model generates an output substantially similar to the input or contains memorized copyrighted works and is made available for others to download, as with open-source and open-weight models.

A critical point often overlooked in GenAI debates is that the Copyright Act of 1976 was drafted with human capabilities and limitations in mind, not AI. The law is structured to account for our human limitations, including imperfect memories and the limited speed at which we can process what we hear, read, or see.[16] As Professor Robert Brauneis explains, the law purposely and logically treats reproduction, derivatives, and distribution differently from the way it treats public displays and public performances.[17] If the argument for GenAI is that its viewing of works is more akin to viewing public displays than making copies, for example, that is misleading.

[15] 17 U.S. Code § 106

[16] Zheng, Jieyu et al., *The Unbearable Slowness of Being: Why Do We Live at 10 bits/s?*, NEURON, https://www.cell.com/neuron/abstract/S0896-6273(24)00808-0 (While streaming a high definition video takes about five million bits per second, or 5 Mbps, and GenAI entities have connections of 300 Gbps, or 30,0000 Mbps, "The information throughput of a human brain is about 10 bits/s.")

[17] Robert Brauneis, *Copyright and the Training of Human Authors and Generative Machines* at 46-51, https://scholarship.law.gwu.edu/cgi/viewcontent.cgi?article=3005&context=faculty_publications

The analogy between AI "learning" and human learning breaks down under scrutiny. When humans view a public display, we don't memorize every detail of it, even when we want to. Even the most dedicated book reader can't comprehend and recall most aspects of more than a few novels a day, and probably cannot retain the same focus and ability to recall information if they read more than a few books a day for more than a week or so.[18] In stark contrast, GenAI can regurgitate quotes verbatim, recall minute details, and consume many thousands of books a day. Indeed, if GenAI could not consume works so rapidly and with such accuracy, the technology would not be economically feasible to produce.[19]

This analysis focuses on the reproduction of copyrighted works for training purposes, rather than on the model outputs.

## IV. Fair Use

Fair use represents one of the most important—and most contested—doctrines in copyright law. Fair use is an affirmative defense, which means that the accused party essentially says, "Yes, we took and used your copyrighted works without your permission and without paying, but we believe we should be allowed to do so because our use is sufficiently different from yours and it enhances or increases scientific progress and the arts without unfairly harming you in the marketplace.

Section 107 of the Copyright Act identifies four factors that courts must consider when assessing whether the use of copyrighted material constitutes infringement or fair use:

1. The purpose and character of the use, including whether such use is of a commercial nature or is for nonprofit educational purposes;
2. The nature of the copyrighted work;
3. The amount and substantiality of the portion used in relation to the copyrighted work as a whole; and
4. The effect of the use upon the potential market for or value of the copyrighted work.

---

[18] *See* Julie Beck, *Why We Forget Most of the Books We Read*, THE ATLANTIC (Jan 26, 2018), https://www.theatlantic.com/science/archive/2018/01/what-was-this-article-about-again/551603/

[19] Even with these capabilities GenAI is not profitable. *See, e.g., OpenAI does not expect to be cash-flow positive until 2029, Bloomberg News reports*, REUTERS (Mar. 26, 2025), https://www.reuters.com/technology/artificial-intelligence/openai-does-not-expect-be-cash-flow-positive-until-2029-bloomberg-news-reports-2025-03-26/

Notably, fair use is a limited exemption, even for noncommercial educational purposes.[20] When examining a fair use claim, courts must explore all four factors and weigh the results together, with due consideration for copyright law's goal of supporting the constitutional mandate to "promote the progress of science and useful arts." While still relevant, factors two and three generally carry less weight than factors one and four.[21] Adding complexity to the analysis, the factors often interact with one another.

### *A. Applying the Four Factors*

In brief, here is how the courts have interpreted each of the factors:

#### *1. Factor One: Purpose and Character of Use*

Factor one asks "whether the new work merely 'supersede[s] the objects' of the original creation, or instead adds something new, with a further purpose or different character, altering the first with new expression, meaning, or message; it asks, in other words, whether and to what extent the new work is 'transformative.'"[22] In *Campbell v. Acuff-Rose Music*,[23] the Supreme Court established that:

> Although such transformative use is not absolutely necessary for a finding of fair use, the goal of copyright, to promote science and the arts, is generally furthered by the creation of transformative

---

[20] *See, e.g.,* Cambridge University Press v. Becker, 446 F. Supp. 3d 1145 (2020). (the copying and distribution 37 excerpts from some books to students was fair use, but that the copying and distribution of 11 excerpts from other books was not fair use.)

[21] *Authors Guild v. Google, Inc.,* 804 F.3d 202, 213–14 (2d Cir. 2015).

[22] *Campbell v. Acuff-Rose Music*, 510 U.S. at 579 (citations omitted) (quoting Folsom, 9 F. Cas. at 348). Importantly, "transform" and "transformative" are different. Transforming web pages into a PDF, making the PDF plain text, and turning that plain text into numbers is very likely not transformative. *See, e.g., Disney Enterprises, Inc. v. VidAngel, Inc*., 869 F.3d 848, 861-63 (9th Cir. 2017) (rejecting argument that encoding of motion pictures to operate a streaming service was a transformative fair use); *Hachette Book Grp. v. Internet Archive*, 664 F. Supp. 3d 370, 380 & n.8 (S.D.N.Y. 2023) (digitizing books not transformative for purposes of fair use); *UMG Recordings, Inc. v. MP3.com*, Inc., 92 F. Supp. 2d 349, 351 (S.D.N.Y. 2000) (same for music); U.S. Copyright Office, Exemption to Prohibition on Circumvention of Copyright Protection Systems for Access Control Technologies, 80 Fed. Reg. 65944, 65960 (Oct. 28, 2015) ("U.S. Copyright Office, Exemption Rulemaking") (rejecting the notion that format-shifting or space-shifting constitutes a fair use); *Columbia Picture Industries, Inc. v. Fung*, 710 F.3d 1020 (9th Cir. 2013) (breaking content into fragments does not make the use non-infringing).

[23] *Id*. at 569.

> works. Such works thus lie at the heart of the fair use doctrine's guarantee of breathing space within the confines of copyright, and the more transformative the new work, the less will be the significance of other factors, like commercialism, that may weigh against a finding of fair use.

*2. Factor Two: Nature of the Copyrighted Work*

Factor two "calls for recognition that some works are closer to the core of intended copyright protection than others, with the consequence that fair use is more difficult to establish when the former works are copied."[24] Notably, "[t]he second factor has rarely played a significant role in the determination of a fair use dispute."[25]

*3. Factor Three: Amount and Substantiality*

Factor three examines whether "'the quantity and value of the materials used,' are reasonable in relation to the purpose of the copying."[26]

*4. Factor Four: Market Harm*

Factor four requires the court to assess "'the effect of the use upon the potential market for or value of the copyrighted work.' It requires courts to consider not only the extent of market harm caused by the particular actions of the alleged infringer but also 'whether unrestricted and widespread conduct of the sort engaged in by the defendant ... would result in a substantially adverse impact on the potential market' for the original."[27]

### *B. Canonical Examples of Fair Use*

This section briefly describes a few canonical examples of courts finding fair use frequently cited in litigation and policy briefs. They are included here to ground the discussion of fair use in more practical terms and illuminate how the court has reasoned through prior cases.

---

[24] *Id.* at 586.

[25] 804 F.3d at 220. Note also that unpublished works, like unpublished manuscripts or movies, are less likely to be fair use, and by scraping most of the public internet, it is possible GenAI companies (unintentionally) collect such works.

[26] 510 U.S. at 586 (quoting Folsom, 9 F. Cas. at 348).

[27] *Id.* at 579 (citations omitted).

### *1. Parody of a Song or Book*

In *Campbell v. Acuff-Rose Music, Inc.,* the copyright holder of a song sued a rap group for copyright infringement. The original song "Oh, Pretty Woman" and the subsequent parody "Pretty Woman" were not exact copies. The court determined that "the copying [of the first line of the original song] was not excessive in relation to the song's parodic purpose."[28] Furthermore, there was a reasonable perception of the song "commenting on the original or criticizing it, to some degree," that met the threshold for parody.[29] The court also rejected any presumption of invalidity based solely on the parody's commercial success, noting that "the mere fact that a use is educational and not for profit does not insulate it from a finding of infringement, any more than the commercial character of a use bars a finding of fairness."[30] A key element of the court's fair use analysis is to distinguish between the "supress[ion of] demand [and] copyright infringement [, which] usurps it."[31]

Similarly, in *Suntrust Bank v. Houghton Mifflin Co.*, the court analyzed a parody dispute between the famous novel "Gone with the Wind" and the secondary work "The Wind Done Gone." The parody book copied significant elements from the original work, including numerous characters, settings, and plot twists. However, like the parody song in *Campbell*, the parody book's for-profit status was "strongly overshadowed and outweighed in view of its highly transformative use of [the original's] copyrighted elements."[32] Since parody involves core components of free speech, the parody book's critical perspective on the original work weighed strongly in favor of finding sufficient transformative purpose. The parody transformed the original work from a third-person perspective into one that tells a new, critical story from the first-person.

Importantly, the court rejected any suggestion that fame creates broader fair use rights, noting that "it is fundamentally at odds with the scheme of copyright to accord lesser rights in those works that are of greatest importance to the public… To propose that fair use be imposed whenever the social value of dissemination outweighs any detriment to the artist, would be to propose depriving copyright owners of their right in the property precisely when they encounter those users who could afford to pay for it."[33]

---

[28] *Id.* at 570.
[29] *Id.* at 583.
[30] *Id.* at 584.
[31] *Id*. at 592 (quoting Fisher v. Dees, 794 F.2d, at 438).
[32] Suntrust Bank v. Houghton Mifflin Co., 268 F.3d 1257, 1269 (11th Cir. 2001).
[33] *Id.* at 1271 (quoting *Harper & Row* at 559, 105 S.Ct. at 2229–30).

### 2. *Thumbnail Images*

Courts have also found fair use in cases involving thumbnail images for web search results, where even exact copying may be transformative if the copy serves a different function than the original work. Key factors supporting these decisions included the fact that the thumbnails were not a substitute for the original images and that they actually helped direct users to the source.

In *Perfect 10 v. Amazon.com,* the court found that:

> "[A] search engine transforms the image into a pointer directing a user to a source of information. Just as a 'parody has an obvious claim to transformative value' because 'it can provide social benefit, by shedding light on an earlier work, and, in the process, creating a new one,' a search engine provides social benefit by incorporating an original work into a new work, namely, an electronic reference tool. Indeed, a search engine may be more transformative than a parody because a search engine provides an entirely new use for the original work, while a parody typically has the same entertainment purpose as the original work."[34]

The court in *Kelly v. Arriba* reached a similar conclusion, finding that thumbnail images were fair use because of the search engine's transformative nature and its benefit to the public (enhancing access to information on the Internet).[35] Additionally, thumbnail images did not harm the photographer's market for the original images and were arguably more transformative than Google's thumbnail use since Google actually superseded Perfect 10's right to sell reduced-size images for cell phones.

### 3. *Google Books*

Perhaps the most frequently cited fair use case in GenAI litigation involves Google's wholesale copying of books to create a book search feature. In *Authors Guild v. Google, Inc.*, the court emphasized that Google's search feature "adds important value to the basic transformative search function, which tells only whether and how often the searched term appears in the book. Merely knowing that a term of interest appears in a book does not necessarily tell the searcher whether she needs to obtain the book because it does not reveal whether the term is discussed in a manner or context falling within the scope of the searcher's interest."[36]

[34] *Perfect 10 v. Amazon.com*, 508 F.3d 1146, 1165 (9th Cir. 2007) (quoting Campbell, 510 U.S. at 579, 114 S.Ct. 1164).

[35] *Kelly v. Arriba*, 336 F.3d 811 (9th Cir. 2003).

Notably, Google implemented significant technical limitations to prevent market substitution. When Google surfaced small verbatim copies of the text, the results were "arbitrarily and uniformly divided by lines of text, and not by complete sentences, paragraphs, or any measure dictated by content, a searcher would have great difficulty constructing a search so as to provide any extensive information about the book's use of that term."[37] This meant that:

> "Google has constructed the snippet feature in a manner that substantially protects against its serving as an effectively competing substitute for Plaintiffs' books. In the Background section of this opinion, we describe a variety of limitations Google imposes on the snippet function. These include the small size of the snippets (normally one eighth of a page), the blacklisting of one snippet per page and of one page in every ten, the fact that no more than three snippets are shown—and no more than one per page—for each term searched, and the fact that the same snippets are shown for a searched term no matter how many times, or from how many different computers, the term is searched. In addition, Google does not provide snippet view for types of books, such as dictionaries and cookbooks, for which viewing a small segment is likely to satisfy the searcher's need."[38]

Related findings included that "[s]nippet view, at best and after a large commitment of manpower, produces discontinuous, tiny fragments, amounting in the aggregate to no more than 16% of a book. This does not threaten the rights holders with any significant harm to the value of their copyrights or diminish their harvest of copyright revenue."[39]

It also seemed important that Google Books provides a link to the source. "A brief description of each book, entitled "About the Book," gives some rudimentary additional information, including a list of the words and terms that appear most frequently in the book. It sometimes provides links to buy the book online and identifies libraries where it can be found."[40]

[36] 804 F.3d at 217-18 (2d Cir. 2015).
[37] *Id.* at 222-23.
[38] *Id.* at 222.
[39] *Id.* at 224.

Finally, although mentioned only in a footnote, the court noted that "Google now honors requests to remove books from snippet view."[41]

### 4. *Critical Analysis*

These canonical cases reveal important limitations on fair use that are often overlooked in debates about GenAI. First, none of the cases involved GenAI, so any argument relying on analogies to GenAI and these cases must acknowledge the inherent limitations. The technology underlying GenAI is meaningfully different from the technology at issue in each canonical case in several significant ways, including the scope and scale of the data required, model size, and compute requirements. More importantly, each technology at issue in those cases (books, a songs, thumbnail images, Google Books) served a discrete, limited function and were therefore inherently more constrained and less potentially disruptive than GenAI models, which are explicitly designed to be general-purpose tools that may disrupt entire industries and fields of employment, such as customer service, visual arts, marketing, and software development.

Perhaps most significantly, in the cases of thumbnails and Google Books (among the cases most frequently cited by defenders of GenAI), the courts focused on the functional role of the technology (e.g., facilitating search), rather than the exploitation of the author's *expressive* content. GenAI, in sharp contrast, is wholly dependent on exploiting expressive content because the expressions are directly encoded into the model itself. This is precisely how the model can quote and regurgitate works, generating "in the style of" outputs. Moreover, Google Books and image search will work the same regardless of whether the book is full of random letter groupings or the image is the visual equivalent of white noise.

---

[40] *Id.* at 209. Here are some other important considerations: First, deleting data from datasets is only an effective solution if no training has commenced on the data. Otherwise, the model developers retain the benefit of the data in the model. Second, it is unclear what is being deleted (if anything): just PDFs/jpegs? Plain text extracted from PDFs and jpegs? The tokens created from the plain text? The token IDs assigned to each token? Third, the onus is usually on the content creators to specifically identify what must be deleted, and non-exact matches likely won't be deleted. Finally, there is usually no confirmation of what was deleted, if anything, and there is no way to verify that the deletion occurred. Unsurprisingly, OpenAI is in no rush to make opting out easier and more transparent. *See* Kyle Wiggers, *OpenAI Failed to Deliver the Opt-Out Tool it Promised by 2025*, TECHCRUNCH (Jan 1, 2025), https://techcrunch.com/2025/01/01/openai-failed-to-deliver-the-opt-out-tool-it-promised-by-2025/

[41] *Id.* at footnote 2.

Those functional technologies are content agnostic. In contrast, GenAI would be worthless if it were similarly ambivalent about the quality of the works it ingests.

GenAI advocates frequently cite several other cases in support of their position that GenAI's use of copyrighted works without authorization constitutes fair use. These include *A.V. v. iParadigms, LLC* ("*iParadigms*"),[42] *Sony Computer Entertainment, Inc. v. Connectix Corp.* ("*Sony Computer*"),[43] *Sega Enterprises Ltd. v. Accolade, Inc.* ("*Sega*"),[44] and *Google, LLC v. Oracle America, Inc.* ("Oracle").[45] However, Charlesworth persuasively and thoroughly notes the distinctions that separate those cases from GenAI.[46] She captures the key differences by observing,

---

[42] 562 F.3d 630 (4th Cir. 2009).

[43] 203 F.3d 596 (9th Cir. 2000).

[44] 977 F.2d 1510 (9th Cir. 1992).

[45] 593 U.S. 1 (2021).

[46] *See, e.g.,* Charlesworth, Jacqueline, *Generative AI's Illusory Case for Fair Use* at 26 for *Kelly*, "The court pointedly distinguished this purpose from copying to capitalize on "artistic expression." It found the search engine's use to be nonsubstitutional because, unlike Kelly's photos, the search and indexing function was "unrelated to any aesthetic purpose."; at 26 for *Perfect 10*, "Invoking *Kelly*, the Ninth Circuit once again held that a search engine's copying of images for thumbnail display was a transformative fair use because the images were not being used for their intrinsic purpose but to create "an electronic reference tool."; at 26 for *Google Books* and *HathiTrust,* "In both *Google Books* and *HathiTrust*, the court established a dividing line between uses that were functional and nonsubstitutional in nature, and uses that were not. Despite its superficial similarity to the mass copying in *Google Books* and *HathiTrust*, then, AI copying cannot be squared with the fair use finding in either of these cases, the holdings of which were careful to preserve copyright owners' legitimate interest in the expressive content of their works."; at 27 for *iParadigms*, "The fact that iParadigms was not seeking to exploit student essay content as such readily distinguishes the copying at issue in iParadigms from that engaged in by generative AI systems."; at 27 for *Sega*, *Sony Computer*, and *Oracle* Charlesworth notes that "the copying was undertaken with respect to a specific work to facilitate interoperability—clearly not the objective of AI copying." For *Sega,* "Accolade's works did not incorporate any of Sega's creative expression. The intermediate copying in Sega cannot be equated with copying by AI systems, the very different purpose of which is to capture and use expressive content."; at 28 for *Sony Computer*, "the holding was premised on the court's determination that the code at issue "contain[ed] unprotected functional elements" that could not be accessed or studied without copying." "As its name would suggest, intermediate copying does not involve ongoing engagement with or exploitation of protected content. The defendants in *Sega* and *Sony Computer* were not seeking to replicate or profit from the plaintiffs' artistic works but instead to produce independently created, compatible products. An AI model, by contrast, is designed permanently to exploit copied creative expression."; at 28 for *Oracle*, "the Court was focused on the interoperability of human programmers."; Finally, at 31 for addressing the argument that GenAI makes transformative use of copyrighted works, she states that "Copying of protected works by generative AI systems has no similar claim to transformativeness. Works are copied in their entirety and mechanically encoded in the AI model without offering any search functionality or other utility to users, let alone criticism or commentary. They are copied to capture their

"In these earlier cases, the determination of fair use turned on the fact that the alleged infringer was not seeking to capitalize on expressive content—exactly the opposite of generative AI."[47]

## V. Why GenAI's Use Might Be Fair Use

There are several arguments for how GenAI might satisfy the fair use criteria detailed in the statute and case law. GenAI companies have shared their stances in litigation, interviews, and in comments to the government. For example, OpenAI submitted a comment to the US Patent and Trademark Office declaring that its use of copyrighted works is fair use.[48] Like all prominent arguments in favor of fair use for GenAI, they lean heavily on transformation under Factor 1.

Here is a credulous overview of how the arguments tend to apply:

> Factor 1: The use of copyrighted works is transformative. The input is copyrighted content, but the output is entirely different, driven by statistical patterns recognized by the GenAI rather than the mere compression of data or the stitching together of data sources. Models are not like databases, storing copies of the original. This allows the models to perform such actions as generating a song about living in the ghetto in the style of Johnny Cash. Moreover, while the purpose of most copyrighted content is to provide entertainment, GenAI uses the content for an entirely different purpose, and the models are incapable of deriving any enjoyment from the content they consume.
>
> Factor 2: The models are only interested in the underlying facts and ideas of the copyrighted works, not the expressive elements. GenAI models are not designed to capture the expressive elements of copyrighted works. Rather, they aim to understand the underlying relationships between data, such as words and sentences, colors and shapes, sounds and rhythms, etc.
>
> Factor 3: Enormous datasets are necessary to train state-of-the-art models, and it is only possible to have sufficiently large datasets by including copyrighted material. There is no getting around the fact that training high-

---

expressive value."

[47] Charlesworth, Jacqueline, *Generative AI's Illusory Case for Fair Use* at 1.

[48] United States Patent and Trademark Office Department of Commerce, Comment Regarding Request for Comments on Intellectual Property Protection for Artificial Intelligence Innovation (2019), https://www.uspto.gov/sites/default/files/documents/OpenAI_RFC-84-FR-58141.pdf

> quality GenAI requires training on entire copies of copyrighted works. The effectiveness of GenAI relies, in part, on the breadth of data. However, the copies are kept from the public. As long as the copies are not revealed to the public, courts should not consider this Factor weighing against the company.
>
> Factor 4: GenAI models do not unfairly compete with copyright owners. Though a model may be trained on a J.K. Rowling book, the model cannot generate a J.K. Rowling-quality novel in the style of J.K. Rowling as an output. Even if the model *could* generate such an output, it would not constitute an infringement. Humans read Rowling's books all the time and then produce novels. Sometimes, Rowling's work even inspires the novels. However, unless the output is substantially similar to the original copyrighted work, there is no infringement, which is the case with virtually all GenAI outputs. The question is not whether the output competes with the original but whether it does so by infringing on the original.

### *A. Necessary Intermediary Copies*

As an example of the arguments in practice, look no further than *UMG Recordings, Inc. v. Suno, Inc.*, and *UMG Recordings, Inc. v. Uncharted Labs (d/b/a Udio)*.[49] Suno and Udio are GenAI companies that trained their models on songs and generate musical outputs. They are represented by the same law firm, and their replies to plaintiffs are similar.

One angle pursued by the GenAI companies was to argue that, separate from the four-factor analysis, it is also not an infringement to make an intermediary copy of a copyrighted work if the copying is necessary to create a non-infringing output:

> …Under longstanding precedent, it is fair use to make a copy of a protected work as part of a back-end technological process, invisible to the public, in the service of creating an ultimately non-infringing new product. Congress enacted the first copyright law in this country in 1791. In the 233 years since, no case has ever—not one single time—reached a contrary conclusion. Each time the question has been presented—and it has been presented over and over and over again—the ultimate

[49] *UMG Recordings, Inc. v. Suno, Inc.*, 1:24-cv-11611, (D. Mass. 2025) and *UMG Recordings, Inc. v. Uncharted Labs*, Inc., 1:24-cv-04777, (S.D.N.Y. 2025).

> conclusion has been that making an "intermediate" copy of a protected work, in the service of generating noninfringing outputs, is permissible, not actionable. *See, e.g., Authors Guild v. Google, Inc.*, 804 F.3d 202 (2d Cir. 2015) (fair use to copy all of the books in numerous university libraries in order to create a commercial, full-text searchable index of the assembled corpus); *Kelly v. Arriba Soft Corp.*, 336 F.3d 811 (9th Cir. 2003) (fair use to copy essentially all of the images on the open internet and show thumbnail versions to users, in the service of creating image-search functionality); *A.V. ex rel. Vanderhye v. iParadigms, LLC*, 562 F.3d 630 (4th Cir. 2009) (fair use to copy student papers into a plagiarism detection tool). The outcome has been no different when the copying has been done in the service of creating an ultimate output that competes with the plaintiff copyright owner's own product. *See, e.g., Sega Enters. Ltd. V. Accolade, Inc.*, 977 F.2d 1510 (9th Cir. 1992) (fair use to copy copyrighted operating system to create unauthorized but non-infringing video game in direct competition with proprietor's own games); *Google LLC v. Oracle Am., Inc*., 593 U.S. 1 (2021) (fair use to copy protected aspects of copyrighted computer software to create a directly competing product).

### ***B. GenAI as Human Learners***

Finally, Suno and Udio also make the common argument that the way GenAI "learns" is little different from how humans learn, implying that if we allow humans to learn in a certain way, then GenAI should be allowed to do so as well.[50]

---

[50] *See also,* Nvidia's response to being sued for scraping YouTube to train its GenAI: "We respect the rights of all content creators and are confident that we are working in full compliance with the letter and the spirit of the law."… "*Anyone is free to learn facts and ideas from publicly available sources*. Creating new and transformative works is not only fair and just, but exactly what our legal system encourages." (emphasis added) Annelise Gilbert, *Nvidia Illegally Scraped YouTube Videos for AI, Suit Says*, BLOOMBERG LAW (August 15, 2024) https://news.bloomberglaw.com/ip-law/nvidia-illegally-scrapped-youtube-videos-to-train-ai-suit-says; *See also* this post by prominent AI researcher Andrew Ng: "just as humans are allowed to read documents on the open internet, learn from them, and synthesize brand new ideas, AI should be allowed to do so too. I would like to see training on the public internet covered under fair use" (2023) https://www.linkedin.com/posts/andrewyng_gpt-4-tells-lies-microscopes-recognize-cancer-activity-7149910997586038784-BP4i

> Like a human musician, Suno did not develop its capabilities in a vacuum. It is the product of extensive analysis and study of the building blocks of music: what various genres and styles sound like; how songs in those genres and styles are harmonized and structured; the characteristic timbres of the instruments and vocalizations in those genres and styles; and so on.

## VI. What About Humans?

Few (if any) scholars argue that GenAI has no colorable argument in favor of fair use. But what fair use advocates take great pains and engage in logical contortions to reach is that somehow the arguments for permitting GenAI companies to take and use any copyrighted works they want are obviously and significantly different from the arguments a human could make to perform the same actions of downloading every movie, song, or book they desire without authorization or payment to the copyright owner.

The arguments for allowing humans a get-out-of-infringement-free card can be strikingly similar to those made for GenAI if we assume the claims made by GenAI advocates are accurate. This becomes clearer as we walk through the same four-factor analysis from the section above:

> Factor 1: The use of copyrighted works is transformative. The input is copyrighted content, but the output is entirely different, driven by patterns recognized by brain neurons rather than the mere compression of data or the stitching together of data sources.[51] Brains are not like databases, storing copies of the original. This allows brains to perform such actions as creating a song about living in the ghetto in the style of Johnny Cash.[52]
>
> Factor 2: Both models and brains are interested in the underlying facts and ideas of copyrighted works *as well as* the expressive elements. Neither

[51] Indeed, it's impossible to understand how to differentiate the logic Judge Alsup applied to LLMs from how the same logic would apply to human brains: "In short, the purpose and character of using copyrighted works to train LLMs to generate new text was quintessentially transformative." Unless a human is only reading a text to reproduce it verbatim, their use would seem to qualify as "quintessentially transformative."

[52] Judge Chhabria notes that Factor One favors fair use because with Meta's Llama model, "Users can ask Llama to edit an email they have written, translate an excerpt from or into a foreign language, write a skit based on a hypothetical scenario, or do any number of other tasks." But I think you'd be hard-pressed to find humans incapable of the same actions if trained to do them.

GenAI models nor brains were specifically designed to capture expressive elements of copyrighted works. Rather, both models and brains aim to understand the underlying relationships between words and sentences, colors and shapes, sounds and rhythms, etc., and they store some representation of them.

Factor 3: Humans do not require mass copyright infringement to be useful, and that should count in favor of humans just as this factor has always favored *less* unauthorized reproduction of copyrighted works, not *more*. More and better books leading to more and better learning applies at least as well to humans as to GenAI training.[53] Humans would also likely produce more breakthroughs, insights, and discoveries if we did not have to pay for copyrighted works, assuming there is an incentive to do so without the protection of copyright law. It is unclear why GenAI companies should be given the discretion to decide when their unauthorized use of copyrighted works supersedes the rights of copyright holders, but humans are not afforded the same deference. Also, like model training data, the information gleaned by brain neurons while consuming the content is never made available to the public.[54]

Factor 4: Humans rarely unfairly compete with copyright owners and do so no more frequently than GenAI models. Though someone may read a J.K.

[53] It's also unclear why LLMs should be allowed to use an unlimited number of complete books to learn, and it is reasonable under Factor Three, but humans cannot pirate even a single book that they may need. If the argument is that more books lead to better models, that holds for human brains as well. This exposes the absurdity of Judge Chhabria's reasoning: "So feeding a whole book to an LLM does more to train it than would feeding it only half of that book. With this in mind, it was "reasonably necessary" for Meta to "make use of the entirety of the works."

[54] Judge Alsup seems to lean into an argument that because LLMs are worse learners than the average human, they deserve more rights: "For one thing, all agree Anthropic needed billions of words to train any given LLM. If using only books, Anthropic would have needed millions of books per model. If using a set comprising only a small fraction of books and a larger fraction of other texts, Anthropic still would have needed hundreds of thousands of books. Authors contend that because Anthropic showed it could use such smaller sets of books, it surely could have used no books at all — or at least not their books. But Authors forget that 'reasonably necessary' does not mean 'strictly necessary.' Authors do not contest that the volume of text required to train an LLM is monumental. Because using so many works was reasonably necessary, using any one work for actually training LLMs was about as reasonable as the next." Left unaddressed is that though a human may require a certain number of works to achieve a graduate-level education, humans are not allowed to infringe freely.

Rowling book, almost nobody can memorize large chunks of the books, even if they want to, unlike GenAI.[55] This makes humans less likely to create an output that is substantially similar to original copyrighted works, which in turn makes humans less likely to produce infringing outputs.

| **A Comparison of Arguments** | |
|---|---|
| **Why GenAI Entities Should be Permitted to Use Copyrighted Work Without Authorization** | **Why Humans Should be Permitted to Use Copyrighted Work Without Authorization** |
| GenAI can transform the content, so the output rarely resembles the exact input. | Humans are even less able to memorize content, making it rarer still for human creations to resemble exact inputs. |
| Taking books/movies/songs/art as inputs is necessary to improve their abilities. | Taking books/movies/songs/art as inputs is equally necessary to improve human abilities. |
| GenAI outputs rarely compete directly with the creators of the inputs. | Human creations compete even less frequently because of inherent human limitations (much lower volume of inputs and outputs, less memorization, and reduced processing power). |
| GenAI does not usually regurgitate training material. Memorization is a bug, not a feature. | Humans are even less capable of memorizing and regurgitating chunks of books, works of art, and other creative works. Therefore, they are less likely to create something substantially similar and more likely to create something transformative.<br>Moreover, as noted above, some memorization is actually desirable for both model developers and humans, making it disingenuous and incorrect to characterize it as a bug, rather than a feature. |

[55] Indeed, Rowling's enormous fortune strongly implies people paid for her books and the derivatives based on her books, like the *Harry Potter* movies, rather than pirate them.

| | |
|---|---|
| GenAI provides a benefit to society (making people faster or more creative, for example). | Humans create novel insights, scientific theories, artistic genres, and other innovations that advance society even more than GenAI. |

### *A. Necessary Intermediary Copies*

For the sake of argument, let us assume that the legal allowance for necessary intermediary copies applies to the exploitation of expressive works and not merely to facilitate interoperability or other functional roles. What is reading, viewing, or listening to copyrighted works if not a necessary intermediary step to a non-infringing output? It is impossible for a human to read, view, listen to, or otherwise consume a copyrighted work without first copying some aspects of the content into the brain. When a person reads a book, they do not store their insights in an external device during the reading process—the "copying" occurs internally and instantaneously.

Yet, while a dataset stores copyrighted works as zeros and ones, the human brain stores inputs as biochemical signals dispersed across neurons and related cells. Moreover, when data is stored in a dataset, it is preserved as a verbatim copy of the original; it can be infinitely copied and shared, and it can persist indefinitely in that form. GenAI then requires additional copies as it feeds the tokens into the model, where some percentage will be perfectly and permanently memorized. Conversely, the brain is an imperfect vessel, making it far less likely that anything "copied" into it will be shared with others in extensive and accurate detail.[56]

These facts favor humans. Even so, the form and format of the necessary intermediate copying should not matter. The most important consideration for how courts should analyze the use of the works is whether it promotes or hinders the progress of science and useful arts.

While some may quibble about whether it would be necessary for humans to consume any copyrighted content they please without authorization to create something new, the same argument applies to GenAI. No single work among the trillions of tokens a model trains on is strictly necessary to produce a non-infringing output. No single work is essential for a model to function, but the works in aggregate improve GenAI capabilities. The same is true of humans and their capabilities.

[56] This is the basis of the childhood game of telephone and what makes it fun.

### *B. GenAI as Human Learners*

This topic is discussed in more detail below under the substantive arguments, but suffice it to say that anyone who claims a human is free to download every song, movie, book, and more that they can find on the public internet to consume for their benefit without consent from the copyright owner or compensation to the owner and claiming that nobody would complain because it is fair use is sorely mistaken.

## VII. Substantive and Related Arguments

I have identified eight common substantive arguments advanced by GenAI proponents and believe none provide compelling reasons to warp the fair use doctrine or copyright law more broadly to favor GenAI companies.

1. Licensing Materials: GenAI should not be limited to only authorized use of copyrighted works.[57]
2. Expressive v. Underlying Ideas: GenAI merely extracts the underlying meta information from the source content.
3. GenAI As Humans: For some purposes, society should afford GenAI the same legal rights as humans for fair use.[58]
4. GenAI as Non-Human: For other purposes, society should not treat GenAI like a human at all.
5. Only Output Matters: The collection and use of copyrighted work as inputs should not matter for copyright analyses.
6. Quantity of Infringing Output: GenAI's generation of infringing content is significantly less important than its frequency of generation.
7. Enabling Greater Creativity: GenAI provides substantial benefits to society and deserves more permissive treatment under copyright law.[59]

---

[57] *E.g.,* Andreessen Horowitz, Notice of Inquiry on Artificial Intelligence & Copyright (Oct. 30, 2023) (claiming it would be "impossible to identify who the relevant rights holders are, and thus there would be no viable way to get statutory royalties to the proper parties.")

[58] *E.g.,* Katie Presscott, *Microsoft boss urges rethink of copyright laws for AI*, THE TIMES (Oct. 21, 2024), https://www.thetimes.com/business-money/companies/article/microsoft-ceo-copyright-laws-information-ai-pc9hh2c0p ("If I read a set of textbooks and I create new knowledge, is that fair use?" he said.)

[59] Chamber of Progress, *Generate and Create*, https://www.generateandcreate.ai/ ("Generative AI is unleashing new waves of creativity including new forms of written, visual, and audio expression.")

8. Stifling Progress: If fair use is not broadly interpreted and applied to GenAI, it will stifle innovation.

These arguments often conflict with one another, as different stakeholders advance different rationales depending on their strategic needs. I have collected the most prominent arguments and crafted the strongest possible version of each—what philosophers call "steel man" arguments—to ensure fair consideration.

For each argument, I demonstrate that there is an equal and opposite argument for every argument in favor of the permissive use of copyrighted works by GenAI companies. I posit that if the counterarguments are at least as compelling as the arguments, there is no special need to favor GenAI over humans. Furthermore, some counterarguments are objectively stronger than the GenAI arguments.

## A. *Licensing Materials: GenAI should not be limited to only authorized use of copyrighted works.*

### 1. *GenAI Argument*

First, publicly available content should be considered *per se* fair use.[60] The unfettered flow of information is essential to human progress. GenAI, a tool designed to learn and process data, should be afforded the same fundamental right to information. This principle reflects a well-established understanding that individuals who choose to share their creations publicly relinquish a degree of control over how that content is used. This is analogous to a filmmaker incorporating a publicly displayed statue into a film without seeking permission. Just as the statue serves as a contextual element enhancing the film, publicly available data enriches the learning process of AI models. To argue otherwise is to suggest that GenAI, unlike humans, is uniquely prohibited from learning from the vast expanse of publicly accessible knowledge. Society must resist the urge to impose archaic legal frameworks on a technology with the potential to revolutionize society.

Second, fair use improves competition. Restricting AI training solely to licensed materials imposes an insurmountable barrier to entry for startups and emerging competitors. Acquiring comprehensive datasets through traditional licensing is prohibitively complex, time-consuming, and expensive. Requiring proper licensing would effectively grant an oligopoly on advanced GenAI development to a handful of tech giants with the resources to navigate this labyrinthine process. Smaller companies, startups, and non-profit organizations would be severely disadvantaged and unable to compete on a level playing field. The rich would get richer, and the rest would be relegated to the sidelines.

---

[60] Examples of this argument can be found in, *e.g.*, OpenAI, *OpenAI and Journalism*, OPENAI (January 8, 2024), https://openai.com/blog/openai-and-journalism; and Memorandum of Law in Support of OpenAI Defendants' Motion to Dismiss at 12 (February 26, 2024), *The New York Times Company v. Microsoft*, No. 1:23-cv-11195 (S.D.N.Y.).

*2. Counterargument*

Addressing the first part of the argument, 'publicly available' does not necessarily mean 'public domain'. The bedrock principle underlying the free sharing of content is a mutually beneficial exchange. Creators often willingly disseminate their work in exchange for tangible rewards, such as advertising revenue, or intangible benefits, like audience growth, recognition, and reputation building. This delicate equilibrium is disrupted when GenAI consumes content without attribution, compensation, or authorization. Such actions deprive creators of the rightful returns on their intellectual investment and erode the incentive structure that fosters the creation and sharing of valuable content in the first place.

The breadth of legitimate reasons for public content creation demonstrates the complexity of creators' motivations:

| **Reasons Why a Person May Want Their Content to be Public While *Not* Wanting GenAI Bots to Scrape Them** | |
|---|---|
| **To Generate Revenue** | • Subscriptions<br>• Licenses<br>• Pay for premium content<br>• Order merchandise<br>• Pay for services (e.g., consultation)<br>• Affiliate marketing<br>• Ad revenue |
| **To Build a Brand or Reputation** | • To gain the attention of possible employers or an audience<br>• To gain the attention of possible partners<br>• To create a fan base or following<br>• To establish their credibility |
| **To Connect with Visitors** | • To build relationships with like-minded folks<br>• To share interests with fellow hobbyists |
| | • To raise awareness of a topic<br>• To share an opinion and hope to convince others of a position |

It is also intellectually dishonest to say anything publicly available is fair game. Almost anything can be found on the public internet because GenAI employees and hackers can and do take content, such as books or movies, from one website and post it on another, ignoring terms of service and subverting requirements for paywalls and accounts.[61] Saying that models should be able to train on what is publicly available is not meaningfully different from saying models should be able to train on any content they can acquire, which would encompass virtually all non-secret content.[62]

Society must also consider the incentive structure that such a permissive stance would create. If more content makes GenAI better, and better content (longer, thoroughly edited, factually accurate, varied syntax, etc.) improves model performance more than low-quality content,[63] we can be all but certain a market will open for people to take content from behind paywalls and other technologically protective measures and put them on the public internet so GenAI companies can "legally" use them under fair use.

Regarding the second part of the argument, blanket fair use is unnecessary for competition in GenAI. The argument that a licensing scheme is impossible or impractical is premature and without merit. The tech industry has demonstrated an unparalleled capacity for innovation and problem-solving. It is incumbent upon these companies to invest the resources necessary to develop a fair and equitable licensing framework or to make at least a good-faith attempt to do so before making conclusory statements about impossibility. Such a framework, once established, can then be adapted for smaller entities, ensuring a level playing field.

---

[61] *See, e.g.*, Alex Reisner, *Revealed: The Authors Whose Pirated Books Are Powering Generative AI*, THE ATLANTIC (Aug. 19, 2023) https://www.theatlantic.com/technology/archive/2023/08/books3-ai-meta-llama- pirated-books/675063/; Samantha Cole, *Leaked Documents Show Nvidia Scraping 'A Human Lifetime' of Videos Per Day to Train AI*, 404 MEDIA (Aug. 5, 2024) https://www.404media.co/nvidia-ai-scraping-foundational-model- cosmos-project/; Cade Metz, Cecilia Kang, Sheera Frenkel, Stuart A. Thompson, and Nico Grant, *How Tech Giants Cut Corners to Harvest Data for A.I.*, N.Y. TIMES (April 6, 2024) https://www.nytimes.com/2024/04/06/technology/tech-giants-harvest-data-artificial-intelligence.html

[62] Hiding behind "permissively licensed" is similarly problematic. Just because something is released under Apache 2.0, for example, doesn't mean it is public domain or that there are no obligations for those who choose to use artifacts with such licenses. They must still provide the Copyright Management Information, for example, as required by the terms of the license, among other requirements. Ignoring those terms can be either a copyright infringement or a breach of contract, depending on how the content owner chooses to enforce the license.

[63] Shayne Longpre et al., *The Responsible Foundation Model Development Cheatsheet: A Review of Tools & Resources* (2024) at 12 ("In the context of cleaning and filtering web data, high-quality has referred to data known to have been written by humans, and has likely gone through an editing process, leading to the development of quality filters which aim to find data most similar to domains such as books or Wikipedia."), https://arxiv.org/abs/2406.16746

Finally, as Judge Chhabria put it:

> But the suggestion that adverse copyright rulings would stop this technology in its tracks is ridiculous. These products are expected to generate billions, even trillions, of dollars for the companies that are developing them. If using copyrighted works to train the models is as necessary as the companies say, they will figure out a way to compensate copyright holders for it.[64]

### ***B. Expressive v. Underlying Ideas: GenAI merely extracts the underlying meta information from the source content.***

#### *1. GenAI Argument*

It is necessary to distinguish between the protection of ideas and the protection of expression. The material on which models are trained is utilized solely for its objective or factual usefulness, not for reproducing creative expressions. This principle reflects the broader legal understanding that copyright protects the expression of ideas, not the ideas themselves. To illustrate, consider that human authors often draw inspiration from existing works and may produce writings that address similar themes or subjects. As long as these new works do not replicate the original expressions or unique structures of the source material, they do not infringe upon copyright protections.

Models operate within the bounds of these established legal frameworks. They process information based on its objective content and generate outputs that do not substantially replicate any specific work. Therefore, they adhere to the same legal standards that govern human creators, ensuring that the rights of original authors are respected while allowing for the generation of new and original content.

Finally, GenAI is solely interested in the metadata associated with content,

---

[64] Order Denying the Plaintiffs' Motion for Partial Summary Judgment and Granting Meta's Cross-Motion for Partial Summary Judgment, *Kadrey v Meta Platforms, Inc.*, No. 3:23-cv-03417-VC-TSH, 2025 WL 56789, (N.D. Cal. June 25, 2025).

such as the correlation between words and concepts. This focus on statistical patterns, rather than the expressive qualities of the work, exempts their behavior from copyright infringement. GenAI is, in essence, a complex algorithm that makes probabilistic guesses devoid of genuine comprehension.

Here's how Judge Chhabria put it:

> But unlike the uses in those cases, Meta's use of the plaintiffs' books does depend on the books' creative expression. As Meta itself notes, LLMs are trained through learning about "statistical relationships between words and concepts" and collecting "statistical data regarding word order, frequencies [what words are used and how often], grammar, and syntax." Word order, word choice, grammar, and syntax are how people express their ideas. So even though LLMs may only learn about "statistical relationships," those relationships are the product of creative expression. This is true even though, as discussed earlier, Llama consumes that expression in a different way than a person would.[65]

### 2. *Counterargument*[66]

This argument misunderstands both copyright law and the functionality of GenAI. The assertion that GenAI training material focuses solely on informational data, rather than expressive elements, and that these two categories can be neatly and clearly separated, overlooks the intricate interplay between information and expression that is essential to developing sophisticated AI systems. The argument implies that copyright law protects only works of pure entertainment, yet the Constitution empowers Congress to promote science and the useful arts, not merely works of pure entertainment.

Moreover, data curation for GenAI is an inherently selective process. While it incorporates objective or factual information, it also prioritizes expressive elements. The very nature of language, whether in factual reporting or creative writing, is a blend of ideas and their artful presentation. GenAI does not merely generate data; it produces outputs that exhibit unique styles, tones, and even hints of creativity. These qualities cannot be attributed solely to the underlying facts; the expressive character of the training material undeniably influences them through the vocabulary, syntax, and numerous other elements that GenAI entities choose to include and weight more heavily.

---

[65] Order Denying the Plaintiffs' Motion for Partial Summary Judgment and Granting Meta's Cross-Motion for Partial Summary Judgment, *Kadrey v Meta Platforms, Inc.*, No. 3:23-cv-03417-VC-TSH, 2025 WL 56789, (N.D. Cal. June 25, 2025).

[66] There are many instances when the expression and the underlying facts *are* separable. This counterargument only applies to GenAI.

This dual nature of training data has profound implications for copyright analysis. It would be absurd to believe that when a human creates a story inspired by *Harry Potter,* it is because they captured the expressive essence of *Harry Potter* only as expression in their neurons when reading the book, but when a model imitates *Harry Potter,* it did so only relying on statistical meta-associations entirely divorced from expression when consuming the book as an input. To make such a claim, GenAI companies must possess a deeper understanding of neuroscience than neuroscientists, as the mechanisms by which the brain learns to associate different words and portions of words with higher-level concepts remain unknown to the specialists.[67]

Furthermore, GenAI developers consciously curate datasets to optimize GenAI performance. The emphasis on including and giving more weight to sources like the *New York Times* than posts on *X* in datasets and when training models goes beyond mere factual accuracy and reflects recognition of the value placed of expressive quality.[68] A model trained exclusively on less refined sources would produce inferior results even if the underlying facts and ideas were the same. This deliberate selection process underscores the role of expressive content in shaping GenAI capabilities.

Finally, disentangling expressive elements of works from metadata and underlying facts is often impossible. When humans read, they typically internalize and apply concepts rather than memorize verbatim expressive content. When recalling information, most people paraphrase rather than retain the precise expressive elements. Few can recite anything they have consumed, chapter and verse (or even a complete paragraph), like GenAI.[69] Because expressive elements are the protected portions under copyright, and GenAI is more likely to memorize its inputs and output them verbatim, GenAI is more likely to infringe upon copyrighted material than humans.

---

[67] Scientist recently made the most detailed map of a portion of a brain ever: a cubic millimeter of a mouse brain. https://www.nature.com/articles/d41586-025-01088-x. For a GenaI company to claim it knows how a human brain works and to assert with confidence that it's radically different from GenAI in a way that is legally meaningful for copyright law is laughable.

[68] *See* Shayne Longpre et al., *The Responsible Foundation Model Development Cheatsheet: A Review of Tools & Resources* (2024) at 4 ("It is common practice to upweight domains of "high-quality" data; data that is known to be written by humans and has likely gone through an editing process such as Wikipedia and books."), https://arxiv.org/pdf/2404.09479.

[69] As an example, even if someone offered me a billion dollars to recite any paragraph from this paper which I have spent several hours working on, I would be unable to do so. I'd be happy to hallucinate a probable answer, though!

### *C. GenAI As Humans: For some purposes, society should afford GenAI the same legal rights as humans for fair use.*

*1. GenAI Argument*

GenAI is rapidly evolving into sophisticated digital entities capable of learning, understanding, and generating human-quality text. Its capacity to process and analyze vast amounts of information resembles the human cognitive process. As such, society must extend to GenAI the same fundamental rights afforded to humans, particularly the freedom to access and learn from publicly available information. As Judge Alsup wrote, "But Authors' complaint is no different than it would be if they complained that training schoolchildren to write well would result in an explosion of competing works."[70]

Humans are the epitome of lifelong learners. Our brains constantly absorb data from the world around us, from the books we read to the conversations we have. This accumulation of knowledge is essential to our intellectual growth, creativity, and problem-solving abilities. GenAI, in essence, replicates this process on a digital scale. It ingests massive datasets, processes information, and learns to generate human-like text.

As Judge Chhabria notes:

> …LLMs have "end uses" including serving "as a personal tutor," assisting "with creative ideation," and helping users "generate business reports." And several of the plaintiffs testified to using LLMs for various purposes, all distinct from creating or reading an expressive work like a novel or biography—for instance, to find recipes, get tax or medical advice, translate documents, or conduct research. All of these functions are different from the use to which the plaintiffs' books are generally put. So copying the books to develop a tool that can perform those functions is a use with a different purpose and character than the books themselves.[71]

Restricting GenAI's access to even public information is akin to denying humans the right to learn.[72] When humans create something new, we invariably draw inspiration from humanity's collective knowledge. GenAI operates on the same principle. The model training process, whereby GenAI refines its abilities based on datasets, does not diminish the original work of authors or artists. It is analogous to a human being inspired by a book or painting to create something novel.

[70] Order on Fair Use, *Bartz v. Anthropic*, No. C 24-05417 WHA, (N.D. Cal. June 23, 2025).

[71] Order Denying the Plaintiffs' Motion for Partial Summary Judgment and Granting Meta's Cross-Motion for Partial Summary Judgment, *Kadrey v Meta Platforms, Inc.*, No. 3:23-cv-03417-VC-TSH, 2025 WL 56789, (N.D. Cal. June 25, 2025).

*2. Counterargument*

Despite its remarkable capabilities, or, more precisely, because of them, GenAI is not a human mind. Humans are inherently limited in their cognitive capacity. We absorb information, process some of it, and retain fragments, often only fleetingly. Even the most voracious reader cannot match the encyclopedic knowledge instantaneously accessible to GenAI.

While a human might struggle to recall specific details from a book read years ago, GenAI can access minute information from millions of books on command. This is not a mere difference in degree but a chasm in kind. The human brain is a biological system with inherent constraints; GenAI is a computational system operating on a level beyond human comprehension.

The argument that GenAI should enjoy the same freedoms as humans to utilize copyrighted material ignores the distinction between human cognition and machine learning.[73] Humans are naturally selective in their consumption and creation, influenced by personal experiences, biases, and creative sparks. GenAI, conversely, is algorithmic, processing information indiscriminately and at an unprecedented scale.[74]

Most other justifications for a finding of transformativeness for GenAI tend to hinge on how they can perform a variety of tasks. But just because you and I could write a poem about an accounting textbook does not mean we can freely reproduce copies of accounting textbooks. There are no examples of GenAI performing some task that humans could not do if allowed to take and use any copyrighted works they want. Moreover, almost all uses anyone makes of any copyrighted work will have a different purpose and character than the copyrighted work itself. This, again, does not excuse copyright infringement for humans, and it should not excuse it for GenAI companies.

---

[72] *See* Ben Horowitz, Brad Smith, Marc Andreessen, and Satya Nadella, *AI for Startups*, Andreessen Horowitz (Nov 1, 2024), https://a16z.com/ai-for-startups/ (A joint letter from Microsoft and the venture capital firm a16z says, in part, "Copyright law should not be co-opted to imply that machines should be prevented from using data—the foundation of AI—to learn in the same way as people.")

[73] It's also the fallacy of extended analogy.

[74] The indiscriminate nature offers no special protection under copyright law. *Cf. Seuss*, 983 F.3d at 454 (finding no transformative use where defendant ComicMix had no particular need to use Seuss's material for its story).

The speed at which models can produce content is also different and plays an important role in Factor Four analysis. As Judge Chhabria said in response to Judge Alsup's comparison of GenAI to humans:

> But when it comes to market effects, using books to teach children to write is not remotely like using books to create a product that a single individual could employ to generate countless competing works with a miniscule fraction of the time and creativity it would otherwise take. This inapt analogy is not a basis for blowing off the most important factor in the fair use analysis.[75]

Additionally, suggesting that humans could similarly exploit copyrighted works without consequence reflects a misunderstanding of the nature of human creativity. First, unlike GenAI, humans generally do not acquire every copyrighted work we want without paying either directly or indirectly. Second, human creation is a complex interplay of experience, emotion, and original thought. Even when influenced by external sources, the human mind invariably adds a unique perspective, unless it is intentionally trying to copy someone else's work. GenAI, on the other hand, can consistently produce outputs that are strikingly similar to its training data unintentionally, which makes such outputs far more common.

As Princeton AI researchers Arvind Narayanan and Sayash Kapoor explain:

> We have a problem now because [repurposing copyrighted works] are being done (1) in an automated way (2) at a billionfold greater scale (3) by companies that have vastly more power in the market than artists, writers, publishers, etc. Incidentally, these three reasons are also why AI apologists are wrong when claiming that training image generators on art is just like artists taking inspiration from prior works.[76]

---

[75] Order Denying the Plaintiffs' Motion for Partial Summary Judgment and Granting Meta's Cross-Motion for Partial Summary Judgment, *Kadrey v Meta Platforms, Inc.*, No. 3:23-cv-03417-VC-TSH, 2025 WL 56789, (N.D. Cal. June 25, 2025).

[76] Arvind Narayanan and Sayash Kapoor, *Generative AI's End-Run Around Copyright Won't*

### *D. GenAI as Non-Human: For other purposes, society should not treat GenAI like a human at all.*

#### *1. GenAI Argument*

Unlike humans, GenAI lacks the capacity for subjective experience and, therefore, cannot enjoy expression as expression. This distinction renders the application of traditional copyright principles to these artificial entities inappropriate. Oren Bracha, a law professor at UT-Austin, articulates this position:

> [N]on-expressive copies do not infringe for a more fundamental reason and with no need to reach the fair use exemption: they do not fall within the subject matter domain of copyright. Non-expressive copies do not involve any use or access to the protected expression as expression and therefore no copyrightable subject matter is taken or enjoyed by the user. The non-expressive reproduction is a mere physical fact that has nothing to do with copyright whose proper and only domain is expression.[77]

Copyright law is predicated on the protection of human expression. When a person engages with copyrighted work, they do so on a personal level, deriving enjoyment, inspiration, or other subjective responses. GenAI, devoid of such faculties, interacts with data in a purely computational manner. It processes information as numerical patterns without experiencing the underlying creative expression. Consequently, to equate GenAI's consumption of copyrighted material to that of a human is to misunderstand the nature of copyright and artificial intelligence.

Furthermore, the transformative nature of GenAI training radically alters the original content. Human consumption preserves a work's original form, whether it be the written word, visual art, or auditory composition. Conversely, LLMs convert these diverse inputs into a numerical representation, a form that is often

---

*Be Resolved by the Courts*, AI SNAKE OIL (Jan. 22, 2024) https://www.aisnakeoil.com/p/generative-ais-end-run-around-copyright; also *see generally see generally* Austin Kleon, STEAL LIKE AN ARTIST: 10 THINGS NOBODY TOLD YOU ABOUT BEING CREATIVE (Workman Pul'g 2012).

[77] See, Oren Bracha, *Generative AI's Two Information Goods* (unpublished).

unrecognizable as the original. As Judge Chhabria puts it:

> …an LLM's consumption of a book is different than a person's. An LLM ingests text to learn "statistical patterns" of how words are used together in different contexts. It does so by taking a piece of text from its training data, removing a word from that text, predicting what that word will be, and updating its general understanding of language based on whether it was right or wrong—and then repeating this exercise billions or trillions of times with different text. This is not how a human reads a book.[78]

*2. Counterargument*

The premise of this argument is flawed. The copies of copyrighted work used to train GenAI models likely *do* fall within the domain of copyright law.[79] The argument also appears to rest on two problematic assumptions: (a) learning from a work makes any use fair use and (b) GenAI development is entirely automated and humans never see reproduced works when developing GenAI.

If the idea is that learning from a copyrighted work is not infringement, so making a copy of something from which to learn is also not infringement, that is clearly wrong. Humans cannot take a book to a photocopier, make a copy, learn from the book, and have the act of reproducing the work forgiven under fair use. Moreover, humans *do* see at least some copyrighted works as they collect, curate, and preprocess the data to train the models. In fact, examining training data is a recommended best practice.[80]

The argument's definitional problems run deeper. If we define "enjoyment" as deriving joy from consuming a work, the argument essentially claims that GenAI should have *greater* leeway under the law than humans because it is *not* conscious, even though the actions of the GenAI company's leadership and employees are intentional.[81] In other words, being *less* like a human should give it *greater* rights than humans. This logic is precisely backwards. Being more human should afford greater rights under the law because the law exists to benefit humans.

---

[78] Order Denying the Plaintiffs' Motion for Partial Summary Judgment and Granting Meta's Cross-Motion for Partial Summary Judgment, *Kadrey v Meta Platforms, Inc.*, No. 3:23-cv-03417-VC-TSH, 2025 WL 56789, (N.D. Cal. June 25, 2025).

[79] "'Copies' are material objects, other than phonorecords, in which a work is fixed by any method now known or later developed, and from which the work can be perceived, reproduced, or otherwise communicated, either directly or with the aid of a machine or device. The term "copies" includes the material object, other than a phonorecord, in which the work is first fixed. 17 U.S.C. § 101 ("Copies")

[80] Shayne Longpre et al., *The Responsible Foundation Model Development Cheatsheet: A Review of Tools & Resources* (2024) at 11 ("You should always spend a substantial amount of time reading through your dataset, ideally at many stages of the dataset design process."), https://arxiv.org/abs/2406.16746

Moreover, this reliance on enjoyment as a defining issue is problematic because joy is inherently subjective. For instance, when reading a textbook, one might not seek joy but rather aim to grasp the underlying facts and ideas, using the text merely as a means to convey information. Many people share a similar experience when reading poetry or listening to certain music. Courts would be misguided to inquire whether every snippet of copyrighted material a person encountered was enjoyable to that person.[82]

If we redefine enjoyment as merely having a subjective experience of consuming a work regardless of whether it is joyful, then the distinction between infringement and non-infringement seems to rely on consciousness. Specifically, it suggests that entities capable of self-awareness regarding the consumption of content as intended should be evaluated differently under the law. But why should a subjective experience be the determining factor in differentiating copyright infringement between GenAI and humans? How does this distinction align with copyright law's purpose of fostering science and the useful arts? The short answer is that it does not. More importantly, GenAI is a tool created by people for the benefit of its creators. Outsourcing infringement to code or a model does not make it legal. For example, Sam Altman cannot personally photocopy a book without authorization, and he should not be able to subvert the law by having OpenAI do it and then profit from that use.

If we instead consider enjoyment as deriving *value* from a work, the issue becomes clearer. GenAI derives value from the content it processes, just as humans do. The model's capacity to "learn" from a work indicates that it derives value from it. If the copyrighted work did not improve the model's capabilities,

---

[81] *E.g.*, it is no accident the copyrighted material was gathered, that it was fed into a model, that some works were weighted more heavily than others, and that the model can generate outputs based on, or verbatim restating, the source material.

[82] As Brauneis puts it, "If the only thing copyright protected in a work were the aesthetic or hedonic reactions that it produced, or if works used to train generative AI were created for the sole purpose of producing such reactions, then that factual difference could make a legal difference…Adopting a view that anhedonic learning, and only anhedonic learning, is a fair use would unfairly disadvantage human authors for whom aesthetic experience and learning are inextricably intertwined." *Copyright and the Training of Human Authors and Generative Machines* at 20.

GenAI companies would not train the model on the work.[83] It is unclear why the value of capability should not matter more than the value of subjective enjoyment, given that the model capabilities are more likely to undermine, disrupt, and compete with the copyright owners' markets than someone's subjective enjoyment.

Regarding the transformation of content, such as words, into numbers as GenAI does, the changes are arguably *less* significant than the transformation occurring in the human brain, where words are converted into complex electrochemical signals. How the brain functions is more mysterious and complex than how LLMs learn. Yet, there are similarities. While Judge Chhabria emphasizes how LLMs learn by predicting the next token, humans similarly learn by predicting what word will likely come next when reading, and it takes years of practice before we are good at reading.[84] Moreover, human learning is no less irreversible. Notably, copyright law does not distinguish between different forms of transformation when assessing infringement. For example, a concert recording converted into an MP3 file is still subject to copyright restrictions, just as GenAI's transformation of words into numbers should be.[85] The encoded words, sounds, and images in a model are merely a derivative work, or, perhaps, a compilation.[86]

---

[83] This is why GenAI developers want as many expressive training materials as they can collect, longer and more detailed descriptions of items in images, and so on. It appears that the latest breakthroughs in computer vision hinge not just on the quantity of data, but on the quality, including "highly detailed image captions," Matt Deitke et al., *Molmo and PixMo: Open Weights and Open Data for State-of-the-Art Vision-Language Models*, https://arxiv.org/abs/2409.17146. Moreover, all artistic works inherently include the expressions of the artists, such as their choices in shading, positioning, and coloring, among others. To consume a work of an artist is to consume the expression itself.

[84] *E.g.*, if you read a sentence in a children's book that began "The cow jumped over the ___," You would know that the answer is unlikely to be "tank." It might be "pale" or "fence," but it's most likely "moon." But that knowledge isn't inherent. You weren't born knowing the phrase. You learned it from repeated exposure.

[85] "Copyright protection subsists, in accordance with this title, in original works of authorship fixed in any tangible medium of expression, now known or later developed, from which they can be perceived, reproduced or otherwise communicated, either directly or with the aid of a machine or device." 17 U.S.C. § 102(a)

[86] 17 U.S.C. § 106 ("[T]he owner of a copyright … has the exclusive right[] to … prepare derivative works based upon the copyrighted work."). A derivative is defined as "a work based upon one or more preexisting works," including "any … form in which a work may be recast, transformed or adapted." 17 U.S.C. § 101 (definition of "derivative work"). A compilation is defined as "a work formed by the collection and assembling of preexisting materials or of data that are selected, coordinated, or arranged in such a way that the resulting work as a whole constitutes an original work of authorship."

Examining how GenAI and humans learn highlights further distinctions. Both processes involve copying information, whether to a model's weights or a brain's neurons. Interestingly, while the brain's retention is more transient than a model's, humans remain uniquely capable of generating novel insights due to our reasoning, curiosity, and experimentation. Thus, if promoting science and the arts is the goal, granting broader permissions to human activities makes more sense than broadening the rights of GenAI entities.

Ultimately, if the argument is that machines consume content differently from humans, we must question whether this distinction genuinely impacts the promotion of science and the arts. Does allowing models to process data advance these fields more effectively than allowing humans to do so? Conversely, why should we believe that permitting human consumption of copyrighted material hinders science and the arts, but allowing models to do the same thing does not?

### E. *Only Output Matters: The collection and use of copyrighted work as inputs should not matter for copyright analyses.*

#### 1. *GenAI Argument*

The crux of copyright infringement analysis for GenAI should be the output, not the input.[87] This principle is grounded in the premise that GenAI and humans should be treated equitably under the law. Just as humans are not liable for copyright infringement when they read and internalize information from a book, GenAI should not be penalized for processing and incorporating copyrighted material during training. While it is undeniable that GenAI requires copies of copyrighted works, both in their datasets and perhaps in their model weights, these reproductions are essential steps in a transformative process that culminates in original outputs.

Human cognition offers a compelling analogy. When we read, our brains create electrochemical representations of the text. This is essentially a form of copying, yet it is not considered infringing. The key lies in the subsequent use of that information: to generate new thoughts, ideas, and expressions. Similarly, GenAI utilizes processed data to produce novel content.

---

[87] This position has been taken by OpenAI's counsel. *See, e.g., Artificial Intelligence and Intellectual Property: Part I–Interoperability of AI and Copyright Law: Hearing Before the U.S. H. Comm. on the Judiciary, Subcommittee on Courts, Intellectual Property, and the Internet 2* (2023), https://judiciary.house.gov/sites/evo-subsites/republicansjudiciary.house.gov/files/evo-media-document/damle-testimony.pdf (statement of Sy Damle); it is also held by prolific copyright law professor Matthew Sag in *Copyright Safety for Generative AI*, 61 Hous. L. Rev. 295, 307-09 (2023), https://papers.ssrn.com/sol3/papers.cfm?abstract_id=4438593 and *Fairness and Fair Use in Generative AI*, 92 Fordham L. Rev. 1887, 1908 (2024), https://ir.lawnet.fordham.edu/cgi/viewcontent.cgi?article=6078&context=flr

To penalize the copying of works into a GenAI's dataset or internal structure without considering the nature of the output is illogical and counterproductive. The sole determinant of copyright infringement should be whether the GenAI's output exhibits substantial similarity to a protected work's expression.[88] Judge Alsup seems to agree with this line of reasoning: "But Claude created no exact copy, nor any substantial knock-off. Nothing traceable to Authors' works."[89] Judge Alsup goes on to write that "Here, once again, there is no allegation of any traceable connection between the Claude service's outputs and Authors' works. The copying used to train the LLMs underlying Claude was thus especially reasonable."

This approach aligns with the core purpose of copyright law: to protect original expression. By centering the analysis on the final product, we ensure that the law continues to serve its intended function without unduly burdening transformative technologies.

### *2. Counterargument*

The exclusive focus on output in evaluating generative AI infringement creates a dangerous precedent. It would mean that anyone could make copies of anything as long as the subsequent creation after reading, listening, and viewing the copies is not substantially similar to the input. Under this logic, a human could read as many pirated books as they want, so long as none of the human's writing after reading those books is substantially similar.[90] Fortunately, and importantly,

---

[88] *See* Matthew Sag, *Copyright and Copy-Reliant Technology*, 103 Nw. L. Rev. 1607, 1638 (2009) ("[C]opyright law appears to embrace a general concept of expressive substitution. To the extent that communication of original expression to the public is the touchstone of copyright infringement, it follows that copyright liability should not ordinarily be found in circumstances where the use in question is incapable of giving rise to any expressive communication."); James Grimmelmann, *Copyright for Literate Robots*, 101 Iowa L. Rev. 657, 664 ("Verbatim copying of a complete work will be protected as fair use if the copy is used solely as input to a process that does not itself use the works expressively. Or, to put it a little more provocatively, nonexpressive uses do not count as reading. They are not part of the market that copyright cares about, because the author's market consists only of readers." (footnote omitted)).

[89] Order on Fair Use, *Bartz v. Anthropic*, No. C 24-05417 WHA, (N.D. Cal. June 23, 2025).

[90] Judge Chhabria seems to make this point unintentionally. Just replace "Llama" with "human brains" in the following: "To say that Meta's downloading was "piracy" and thus cannot be fair use begs the question because the whole point of fair use analysis is to determine whether a given act of copying was unlawful… The purpose of fair use is to allow new expression that won't substitute for the original work, and whether a given use was made in good or bad faith wouldn't seem to affect the likelihood of that use substituting for the original… that downloading must still be considered in light of its ultimate, highly transformative purpose: training Llama… Because Meta's ultimate

the Ninth Circuit has also shot down this logic twice.[91]

Alternatively, perhaps the argument is that outputs only matter when applied to GenAI, and humans should be held to a higher standard (*i.e.*, must have authorization to use the copyrighted works they want to copy). This would mean granting GenAI systems a privilege denied to humans for an action no less harmful (GenAI output still floods the market and undermines the value of creators) and likely less helpful (humans are the only entity that creates breakthrough innovations and profound insights) than if humans were similarly permitted such leeway.

There is no reason to believe humans are not at least as likely to utilize copyrighted works in transformative, socially beneficial ways as GenAI. Despite this, society acknowledges the fundamental right of creators to control the use of their intellectual property, and we expect copyright owners to be fairly compensated.[92] Without this social contract, creators would have little incentive to share their creations. It is inconceivable to suggest that individuals should be permitted to consume all copyrighted materials without permission or compensation, regardless of their potential to contribute to the public good.

## F. *Quantity of Infringing Output: GenAI's generation of infringing content is significantly less important than its frequency of generation.*

### 1. *GenAI Argument*

GenAI outputs are inherently unpredictable and uncontrollable. They operate probabilistically, meaning their responses are based on statistical patterns learned from vast datasets. This characteristic makes it impossible to guarantee that GenAI will never produce content substantially similar to its training data. Such occurrences, while regrettable, are anomalous and the product of a complex system beyond the developer's direct control.

Imposing liability on GenAI developers for these isolated incidents would conflate accidental output with intentional infringement. It would be akin to holding a book publisher responsible for every instance of plagiarism by a reader.

---

use of the plaintiffs' books was transformative, so too was Meta's downloading of those books."

[91] *Sega*, 977 F.2d at 1517-19 ("intermediate copying of computer object code may infringe the exclusive rights granted to the copyright owner in section 106 of the Copyright Act regardless of whether the end product of the copying infringes those rights."), *Sony Computer*, 203 F.3d at 602-03 ("In *Sega*, we recognized that intermediate copying could constitute infringement even when the end product did not itself contain copyrighted material.")

[92] As noted above in Sec. VII(A)(2), compensation can be direct payment, but it can also be ad revenue, subscriber, building a brand or reputation, and more.

The overwhelming majority of GenAI outputs are original and creative, demonstrating the technology's immense potential for innovation.

Moreover, subjecting GenAI developers to the specter of endless litigation for potential copyright infringement needlessly disrupts commerce. The threat of ruinous legal battles will deter investment, hinder research, and ultimately deprive society of the substantial benefits that GenAI can offer.

A rational approach demands focusing on the technology's overall functionality. If a GenAI model consistently generates infringing content, there may be grounds for concern. However, holding developers liable for isolated incidents, particularly when provoked by malicious or adversarial prompts, imposes an unreasonable and counterproductive burden.

*2. Counterargument*

The notion that GenAI should enjoy a privileged status immune from copyright infringement liability when it infringes is both legally unsound and practically unwise. Such an exemption would create an uneven playing field, unfairly advantaging tech giants at the expense of human creators.

A human author can be held liable for copyright infringement for even a minuscule portion of their overall output.[93] This is a key principle of copyright law. Moreover, any argument that GenAI lacks the intention to infringe is irrelevant; copyright infringement does not require intent.

Furthermore, the assertion that GenAI should be treated similarly to humans in some respects, such as learning from existing works, while simultaneously demanding immunity from infringement liability, amounts to claiming the benefits of human rights without accepting the corresponding responsibilities.

Society cannot afford to allow multi-billion-dollar corporations to exploit the creative works of others without consequence. Such a permissive environment would stifle innovation, erode the value of copyright, and ultimately harm our society's cultural and economic fabric. GenAI must be subject to the same copyright laws as humans. To do otherwise would prioritize corporate profits over individual rights and the integrity of the creative ecosystem.

### *G. Enabling Greater Creativity: GenAI provides substantial benefits to society and deserves more permissive treatment under copyright law.*

[93] Consider all the outputs a typical human makes: notes, doodles, texts, blog posts, papers, social media posts, photos, etc.

#### *1. GenAI Argument*

Generative AI is a catalytic force for artistic innovation, lowering the technical barriers to creative expression and empowering a broader spectrum of individuals to contribute meaningfully to the arts. This democratization of creativity is akin to the automobile revolutionizing transportation. While both technologies can be misused, their potential for positive impact is undeniable.

Just as society has invested in infrastructure to support automotive advancement, we must foster an environment that nurtures GenAI's growth and development. By embracing this technology, we unlock a world of artistic possibilities and enrich our society's cultural fabric.

#### *2. Counterargument*

Even assuming, for the sake of argument, that access to copyrighted materials enhances creativity, does this justify the uncompensated appropriation of those materials by GenAI developers? Such a proposition establishes a precedent where creators can be deprived of the fruits of their labor under the guise of inspiring or benefiting others. This logic would dismantle the very foundation of intellectual property rights, granting free rein to anyone to exploit the work of others without recourse, merely because it *might* lead to more creativity.

### ***H. Stifling Progress: If fair use is not broadly interpreted and applied to GenAI, it will stifle innovation.***

#### *1. GenAI Argument*

GenAI represents a profound frontier in technological advancement, with the potential to revolutionize industries, foster creativity, and enhance human capabilities. Imposing restrictions on its development by severely limiting the application of fair use would be counterproductive, as it would limit human potential and hinder societal progress.

Publicly available data serves as the raw material for GenAI development, enabling it to learn, adapt, and evolve. To deny GenAI access to this knowledge is akin to denying humans the right to learn and develop. Such limitations would create an artificial barrier to technological advancement, favoring a select few and hindering GenAI's potential to benefit society as a whole.

Moreover, limiting the reach of fair use would concentrate control over GenAI development in the hands of a few powerful entities with access to troves of data (*e.g.*, Google, Meta, X) and the billions of dollars needed to license additional data, reinforcing an oligopolistic landscape. A diverse and competitive ecosystem is essential for driving rapid progress and ensuring that AI benefits society as broadly as possible.

We must recognize GenAI as an emerging form of intelligence. By denying it access to the same resources available to humans, we risk hindering its growth and, consequently, limiting the potential for societal advancement.

*2. Counterargument*

While GenAI undoubtedly presents opportunities for innovation, it is far from clear that it is overwhelmingly beneficial. We must distinguish between innovation and progress because not all innovation equates to societal advancement. For instance, while GenAI might rapidly generate new forms of creative output, the quality, originality, and overall cultural contribution of such outputs remain to be determined.[94] To make the point more concrete, it is far from evident that the world is better today because of ChatGPT than it was before ChatGPT's existence. Therefore, it is questionable whether society should trade away copyright protections for the convenience of using a chatbot. The onus is on GenAI companies to demonstrate, for example, that chatbots significantly improve student outcomes from tutoring rather than undermine education by enabling easy cheating and mediocre efforts that a chatbot can enhance.[95]

GenAI is a tool. Unlike humans, it does not develop new theories, like how Darwin made the leap from observation to a theory of evolution or how Einstein's thought experiments led to the theory of relativity. Saying that GenAI could lead to innovation is like saying a more powerful particle collider will help physicists learn more about physics. While technically true, it's not a particularly meaningful insight. The collider itself doesn't discover anything. More importantly, unlike when building CERN, we have no reason to believe that any model dependent on copyrighted material, such as a consumer-facing OpenAI GPT model, for example, will lead to profound innovation. There is no theoretical basis for such a claim; it is just speculation. GenAI is like making a plow faster. But tractors aren't what allow America to have enough food when only 2% of the population is farmers.[96] The real catalyst was the human invention of industrial fertilizer. GenAI cannot make such leaps yet and may never be able to. If you want to have the most transformative outcomes, humans have a stronger argument. *All* major scientific breakthroughs and new genres of art were discovered or created by humans. *None* have been discovered or created by GenAI models.

---

[94] Of course, there are less flattering examples of innovation that was not progress, such as the slave trade. GenAI is not like the slave trade, but claims of innovation inherently equaling progress are equally dubious.

[95] Another example would be to show that the ability to rewrite emails and generate marketing copy and other outputs outweighs the harms of easily producing misinformation, disinformation, and scams. It would also be wise to require GenAI entities to disclose their environmental impact from *all* model training and inference (including for models they didn't make public, which is most models) for the full lifecycle (from minerals and materials necessary to create the GPUs, to the water and electricity necessary to run the GPUs, not just the carbon emissions), to justify a claim for special legal exemptions. *See, e.g.,* Jacob Morrison et al., *Holistically Evaluating the Environmental Impact of Creating Language Models* (2024 under review) https://openreview.net/pdf/09f57c232e7b56840cd95b200d0e74eeea7e5db9.pdf

The models that have proven helpful for discovery so far are specialized in pharmacology, geometry, biology, and materials science, rather than being general-purpose models. Other research indicates that while GenAI may raise the creativity floor of individuals, it places a ceiling on the group.[97] This is unsurprising, given that the models themselves have a ceiling based on their training data and the manner in which they were fine-tuned. Because they are more likely to output material they have seen repeatedly, they tend to generate clichéd ideas.

Regarding market consolidation, antitrust law should handle arguments about oligopolies, not copyright law. If the biggest tech companies have an excessive advantage due to proper copyright law enforcement, the solution is not to weaken copyright protections but to strengthen antitrust enforcement to promote competition.

Moreover, the argument for granting GenAI models unrestricted access to copyrighted material establishes a double standard, permitting GenAI entities privileges not afforded to human creators. Humans are subject to rigorous copyright laws when utilizing the work of others. It is inconsistent and unnecessary to exempt GenAI from these same constraints. Training GenAI models on copyrighted data without proper compensation or authorization undermines the economic foundation that supports the creative industries.

### *I. Conclusion*

There is no apparent substantive reason why for-profit GenAI companies should receive a free pass to use the content of billions of webpages without consent. The primary purpose of copyright law is “[t]o promote the Progress of Science and useful Arts.” Yet it is humans, not GenAI, who have developed *all* novel scientific theories, unprecedented mathematical proofs, new music genres, and original forms of visual art. Importantly, humans have accomplished all this without undermining copyright law or demanding mass exemptions under fair use.

---

[96] https://www.fb.org/newsroom/fast-facts

[97] Geoff Brumfiel, *Research Shows AI Can Boost Creativity for Some, But at a Cost*, NPR (July 12, 2024), https://www.npr.org/2024/07/12/nx-s1-5033988/research-ai-chatbots-creativity-writing

If society is to focus on the purpose of the law and the practical effects of granting or withholding a finding of fair use, and not esoteric technical arguments built on something as unrealistic as separating ideas from expression without retaining *any* expressive elements, or drawing artificial lines around whether and to what extent a human enjoyed a copyrighted work, then we are left with only one conclusion. Allowing GenAI companies to take unlimited unauthorized and uncompensated copies of copyrighted works does not further science or the arts to the same extent that humans do without such allowances.

Moreover, if our goal is merely turbocharging science and the arts in the near term, we would be far better served by allowing the only entity ever known to create such breakthroughs —humans—to download and use copyrighted works so long as the output is not substantially similar to the input: humans. Of course, if fair use is applied too liberally, most copyright law collapses because virtually all non-substantially similar outputs could be considered fair use. This, in turn, would devastate the markets for digital goods because nobody would ever have to pay to see another movie, read another book, or hear another song.

## VIII. SUMMARY

This paper has explained that GenAI companies require billions of copies of copyrighted materials to make highly capable models, and these copies are often unauthorized uses that violate copyright holders' exclusive rights. Furthermore, no compelling arguments for fair use apply exclusively to GenAI that do not also apply to humans who also make copies. Below is a recap of how some of the most common arguments advanced by GenAI entities fall short unless humans are entitled to the same leniency.

- The notion that GenAI can separate underlying ideas from expressions when consuming entire copyrighted works in a way that is legally meaningfully different from how humans gather the underlying ideas is a legal fiction. Moreover, this attempted distinction has no practical bearing on the purpose of copyright law, which is to promote the sciences and useful arts. Regardless of whether a machine uses copyrighted works without paying or humans do, the impact on the copyright holder's market is, at best, the same, and most likely worse when GenAI does it.
- Basing copyright law on whether the work consumed was enjoyed is too subjective to be helpful. Additionally, it ignores that copyright law was not created to prevent enjoyment without authorization but to promote science and the useful arts.

- The fact that something is "publicly available" does *not* mean it is in the public domain or that the content creator is not seeking some benefit that ordinarily occurs when humans view the material. Such benefits do not ordinarily occur when GenAI "views" the material.
- If only the outputs matter, then humans should have even greater leeway under fair use to use copyrighted works because humans are *less* likely to produce substantially similar outputs and are *more* likely to create something novel and useful based on the ideas of copyrighted works, such as innovative art forms, valuable inventions, scientific discoveries, profound business insights, and more.

A particularly troubling issue with arguing that GenAI's use of copyrighted work is fair use is that it would mean, in effect, that the government is mandating that everyone must allow GenAI companies to use their copyrighted works if they are "publicly available" or "permissively licensed." There would be no requirement to permit anyone to opt out. If other claims, like unjust enrichment and breach of contract, are preempted by the Copyright Act, as they often are, then there would be little or no recourse for copyright holders to prevent GenAI entities from copying and training on virtually any copyrighted work.

This paper ends with the same question presented in the introduction: What argument works for GenAI but does not apply at least equally well to humans? To the best of my knowledge, there is no such argument. Therefore, society is left with only three options: (1) grant both GenAI and humans a fair use exemption to virtually all copyrighted works, (2) grant only humans an exemption to virtually all copyrighted works, or (3) exercise common sense and grant neither GenAI nor humans an exemption to virtually all copyrighted works.